\documentstyle[preprint,aps,prl]{revtex}
\input{epsf}
\tightenlines
\begin{document}
\draft
%
%==============================================================================
%
%	Title:
%
\title{Evidence for $\nu_\mu \to \nu_e$ Neutrino Oscillations from LSND}
%
%==============================================================================
%
%	Author list:
%
\author{
C.~Athanassopoulos,$^{11}$ 
L.B.~Auerbach,$^{11}$ 
R.L.~Burman,$^6$ 
D.O.~Caldwell,$^3$ 
E.D.~Church,$^1$
I.~Cohen,$^5$ 
J.B.~Donahue,$^6$ 
A.~Fazely,$^{10}$ 
F.J.~Federspiel,$^6$ 
G.T.~Garvey,$^6$
R.M.~Gunasingha,$^7$ 
R.~Imlay,$^7$ 
K.~Johnston,$^8$ 
H.J.~Kim,$^7$ 
W.C.~Louis,$^6$ 
R.~Majkic,$^{11}$
K.~McIlhany,$^{1}$ 
W.~Metcalf,$^7$ 
G.B.~Mills,$^6$ 
R.A.~Reeder,$^9$ 
V.~Sandberg,$^6$
D.~Smith,$^4$ 
I.~Stancu,$^1$ 
W.~Strossman,$^1$ 
R.~Tayloe,$^6$ 
G.J.~VanDalen,$^1$ 
W.~Vernon,$^2$
N.~Wadia,$^7$ 
J.~Waltz,$^4$ 
D.H.~White,$^6$ 
D.~Works,$^{11}$ 
Y.~Xiao,$^{11}$ 
S.~Yellin$^3$\\
(LSND Collaboration)\\
\mbox{}\\
}
\address{$^1$ University of California, Riverside, CA 92521}
\address{$^2$ University of California, San Diego, CA 92093}
\address{$^3$ University of California, Santa Barbara, CA 93106}
\address{$^4$ Embry Riddle Aeronautical University, Prescott, AZ 86301}
\address{$^5$ Linfield College, McMinnville, OR 97128}
\address{$^6$ Los Alamos National Laboratory, Los Alamos, NM 87545}
\address{$^7$ Louisiana State University, Baton Rouge, LA 70803}
\address{$^8$ Louisiana Tech University, Ruston, LA 71272}
\address{$^9$ University of New Mexico, Albuquerque, NM 87131}
\address{$^{10}$ Southern University, Baton Rouge, LA 70813}
\address{$^{11}$ Temple University, Philadelphia, PA 19122}
%
%==============================================================================
%
%	Date:
%
\date{\today}
\maketitle
%
%==============================================================================
%
%	Abstract:
%
\begin{abstract}
%---------------
A search for $\nu_\mu \to \nu_e$ oscillations has been conducted with the 
LSND apparatus at the Los Alamos Meson Physics Facility. 
Using $\nu_\mu$ from $\pi^+$ decay in flight, the $\nu_e$ appearance is 
detected via the charged-current reaction $\nu_e \, C \to e^- \, X$. 
Two independent analyses observe a total of 40 beam-on high-energy electron 
events (60 $<$ $E_e$ $<$ 200 MeV) consistent with the above signature. 
This number is significantly above the $21.9 \pm 2.1$ events expected from the 
$\nu_e$ contamination in the beam and the beam-off background. 
If interpreted as an oscillation signal, the observed oscillation probability 
of $(2.6 \pm 1.0 \pm 0.5) \times 10^{-3}$ is consistent with the previously 
reported $\bar\nu_\mu \to \bar\nu_e$ oscillation evidence from LSND.
\end{abstract}
\pacs{14.60.Pq, 13.15.+g}
%
%==============================================================================
%
%	1 - Introduction
%

In this letter we describe the results of a search for $\nu_\mu \to \nu_e$ 
oscillations using a $\nu_\mu$ flux from $\pi^+$ decay in flight (DIF). 
The data were taken with the Liquid Scintillator Neutrino Detector (LSND) 
at the Los Alamos Meson Physics Facility (LAMPF). 
The result of a search for $\bar\nu_\mu \to \bar\nu_e$ oscillations, using a 
$\bar\nu_\mu$ flux from $\mu^+$ decay at rest (DAR), has already been reported 
in Ref.~\cite{DAR-LSND}, where an excess of events was interpreted as evidence 
for neutrino oscillations. 
The analysis presented here uses a different component of 
the neutrino beam, a different detection process, and has different backgrounds 
and systematics from the previous DAR result, providing an independent check on 
the existence of neutrino oscillations.
%
%==============================================================================
%
%	2 - Neutrino Source
%

\begin{figure}[h]
\begin{center}
\mbox{
\epsfxsize=0.9\columnwidth \epsfbox[20  20 550 550]{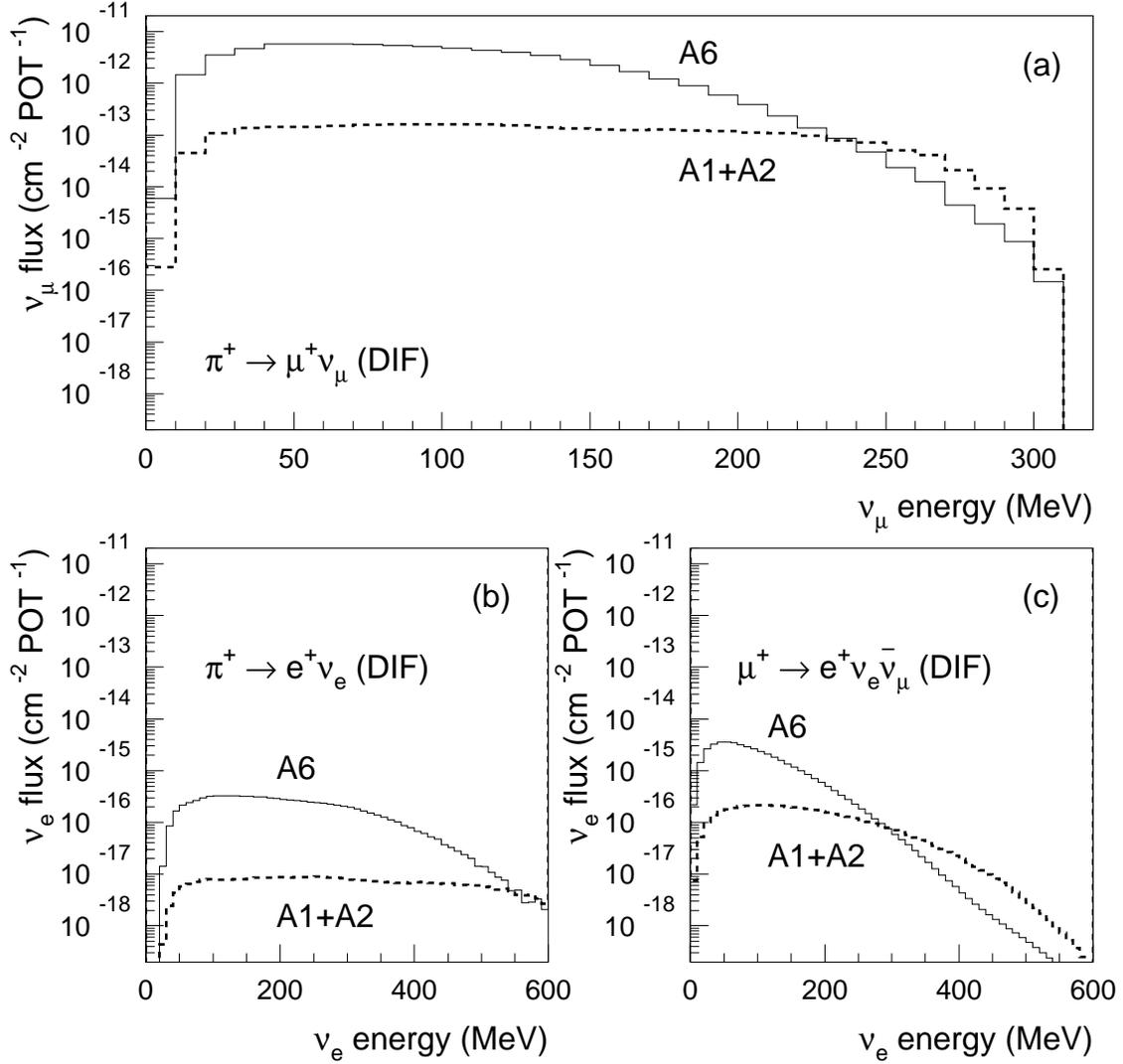}
}
\end{center}
\caption{
Calculated $\nu_\mu$ and $\nu_e$ DIF fluxes at the detector center from the A6 
target (solid histograms) and from the A1+A2 targets (dashed histograms). 
}
\label{Fig1}
\end{figure}

The primary source of DIF $\nu_\mu$ for this experiment is the A6 water target 
of the LAMPF 800 MeV proton linear accelerator. 
Approximately 3.4\%\ of the $\pi^+$ produced in the 30-cm long target decay in 
flight before reaching the water-cooled copper beam stop, situated 1.5 m 
downstream. 
The generated $\nu_\mu$ flux, with energies up to 300 MeV, is illustrated in 
Fig.~\ref{Fig1}(a), as calculated at the center of the detector, 30 m away 
from the beam stop. 
Two upstream thin carbon targets, A1 and A2, located at 135 m and 110 m from 
the detector center, respectively, provide additional small contributions to 
the $\nu_\mu$ flux - also shown in Fig.~\ref{Fig1}(a).
However, for $\nu_\mu \to \nu_e$ oscillations with small $\Delta m^2$ values, 
the $\nu_\mu$ flux from A1 and A2 can have a significant effect due to the 
longer baselines.
The main beam-related backgrounds (BRB) to the $\nu_\mu \to \nu_e$ search 
come from the intrinsic $\nu_e$ component of the beam, shown in 
Figs.~\ref{Fig1}(b) and (c). 
The flux from $\pi^+ \to e^+ \nu_e$ DIF is suppressed by the branching ratio 
of $1.24\times 10^{-4}$, while the flux from $\mu^+ \to e^+ \nu_e \bar\nu_\mu$ 
DIF is suppressed by the longer $\mu$ lifetime and the kinematics of the 
three-body decay. 
The neutrino flux calculations are described in detail in 
Ref.~\cite{Burman-MC} and yield a systematic error of 15\% for the $\nu_\mu$ 
DIF flux. 
The LSND measurement of the exclusive reaction 
$\nu_\mu \, \mbox{}^{12}C \to \mu^- \, \mbox{}^{12}N_{gs}$,
which has a well understood cross section, confirms the calculated flux to 
within a 15\% statistical error~\cite{numuC}.  

The data discussed here correspond to 14772 Coulombs of protons on target 
(POT) during the years 1993 (1787 C), 1994 (5904 C), and 1995 (7081 C). 
The beam duty factor - defined as the ratio of data collected with beam on to 
that with beam off - has a weighted average of 0.07 for the three years.
%
%==============================================================================
%
%	3 - Detector
%

The LSND apparatus, described in detail elsewhere~\cite{NIM-LSND}, consists of 
a steel tank filled with 167 metric tons of liquid scintillator and viewed by 
1220 uniformly spaced $8 ''$ Hamamatsu photomultiplier tubes (PMT). 
The scintillator medium consists of mineral oil $(CH_2)$ with a small admixture 
(0.031 g/l) of butyl-PBD. 
This mixture allows the detection of both \v{C}erenkov and isotropic 
scintillation light, so that the on-line reconstruction software provides 
robust particle identification (PID) for electrons, along with the event 
vertex and direction. 
The electronics and data acquisition (DAQ) systems were designed to detect and 
record related events separated in time.

Despite 2.0 kg/cm$^2$ shielding above the detector tunnel, there remains a 
very large background to the oscillation search due to cosmic rays. 
This background is highly suppressed by a veto shield~\cite{E645}, which 
provides both passive and active shielding. 
It is viewed by 292 uniformly spaced $5 ''$ EMI PMTs and has a threshold 
of 6 PMT hits. 
Above this value a signal holds off the trigger for 15.2 $\mu$s while 
inducing an 18\% dead-time in the DAQ. 
A veto inefficiency $< 10^{-5}$ is achieved off-line with this detector for 
incident charged particles.
The veto inefficiency is obviously much larger for incident cosmic-ray neutrons.

A GEANT-based Monte Carlo (MC) is employed to simulate interactions in the LSND 
tank and the response of the detector system. 
The detector response parameterizations were measured either in a
test beam or in a controlled setting. 
The electron simulation is calibrated below 52.8 MeV using Michel electrons 
from the decay of stopped cosmic-ray muons and then extrapolated into the 
DIF energy range. 
The MC data set used to calculate electron selection efficiencies in the DIF 
analysis (DIF-MC) uses the calculated $\nu_\mu$ flux, 100\% $\nu_\mu \to \nu_e$ 
transmutation, and the $\nu_e \, C \to e^- \, X$ cross section calculated in 
the CRPA model\cite{CRPA}.
%
%==============================================================================
%
%       4 - Event Selection
%

Candidate events for $\nu_\mu \to \nu_e$ oscillation from the DIF $\nu_\mu$
flux consist of a single, isolated electron (from the 
$\nu_e \, C \to e^- \, X$ reaction) in the energy range 60--200 MeV. 
The lower limit is chosen to be well above the endpoint of the Michel electron 
spectrum (52.8 MeV) to avoid backgrounds induced by cosmic-ray muons and 
beam-related $\nu_\mu$ and $\overline{\nu}_\mu$ events. 
The upper limit of 200 MeV is the energy above which the beam-off background 
rates increase, and the expected signal becomes much attenuated. 

A preliminary selection was made to arrive at an initial data sample. 
The electron PID parameters used in the DAR analysis~\cite{DAR-LSND} 
retain high efficiency (98.1$\pm$1.7\%), but have limited background rejection 
in the DIF energy range. 
New PID parameters developed for this analysis are used in final event 
selection as described below. 
To reduce the cosmic-ray muon related background several cuts are made. 
First the veto shield is required to have less than 4 active PMTs. 
Second, the events must be reconstructed more than 35 cm from the surface 
defined by the inner PMT faces.
Third, events with another event of 200--700 hit PMTs in the following 
30 $\mu$s (characteristic for Michel electrons) are dropped if they are 
within 200 cm of the subsequent event, or have more than 600 hit PMTs,
eliminating stopping muons not vetoed by the DAQ. 
Finally, the event must not have had any previous activity above 600 hit PMTs 
or within 200 cm in space in the preceeding 30 $\mu$s, removing remaining 
Michel electrons. 
These cuts have an overall efficiency of 82.4$\pm$2.7\% for electrons in the 
DIF energy range.

The event reconstruction and PID techniques used in the DIF analysis were 
developed to utilize fully the capabilities of the LSND apparatus. 
The basis for the reconstruction is a simple single track event model, 
parametrized by the track starting position and time, direction, and energy. 
For any given event, the expected photon intensities and arrival times are 
calculated from these parameters at all PMTs. 
A likelihood function that relates the {\em measured} PMT charge and time 
values to the {\em predicted} values is used to determine the best possible 
event parameters and also provides PID. 
Two independent analyses were performed, sharing basic goals, but differing 
in approach and parameterizations. 
Both analyses are described in detail elsewhere~\cite{DIF-LSND}.

\begin{figure}[h]
\begin{center}
\mbox{
\epsfxsize=0.9\columnwidth \epsfbox[20  20 550 550]{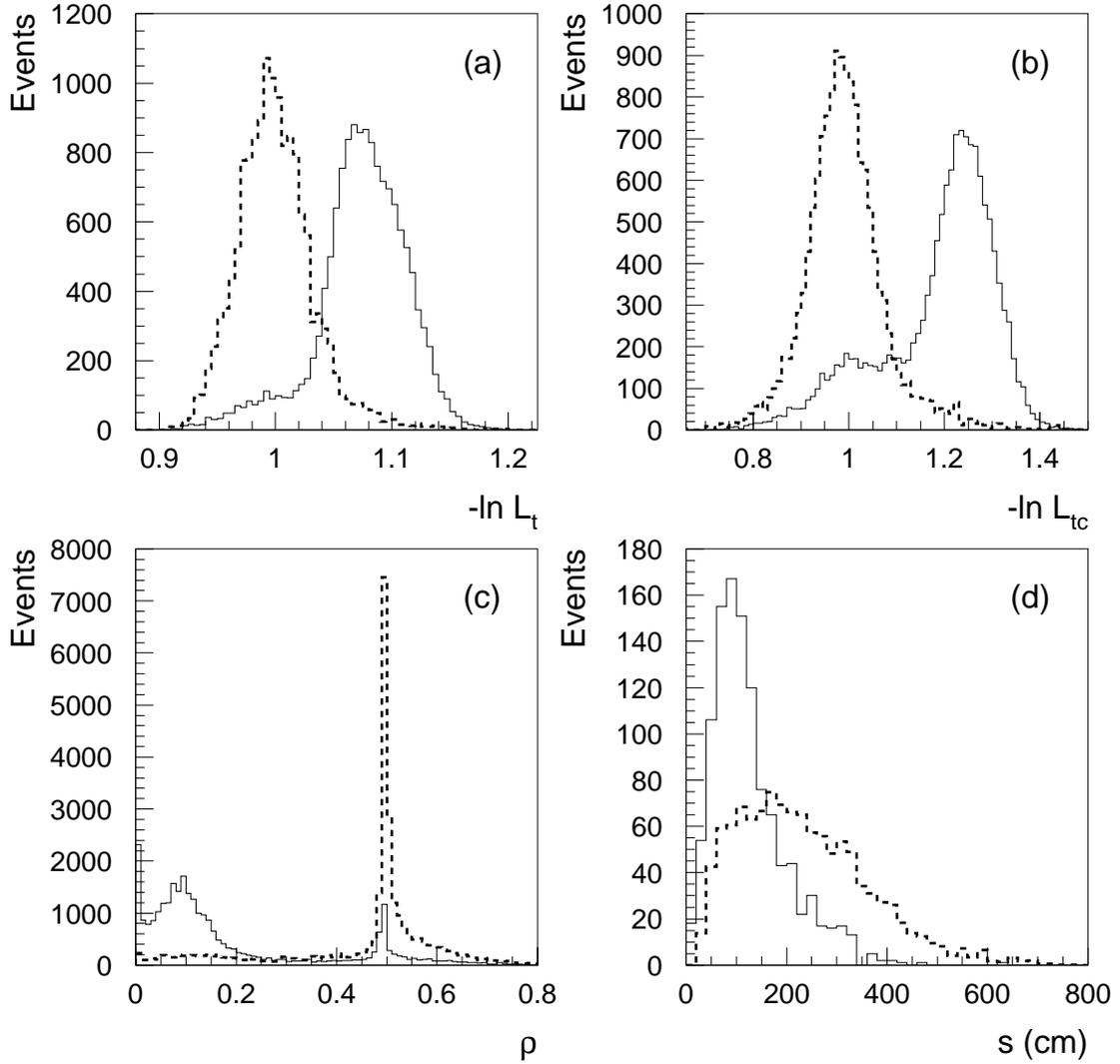}
}
\end{center}
\caption{
Timing likelihoods for (a) the entire event and (b) the \v{C}erenkov region 
only. 
(c) \v{C}erenkov-to-scintillation density ratio, $\rho$. 
(d) Projected track-length to the tank wall intersection. 
(a)-(c) correspond to all (beam on+off) DIF data after the 
pre-selection and (d) after all other cuts have been applied. 
All superimposed distributions (dashed) correspond to the DIF-MC 
simulation, normalized to the same areas.
}
\label{Fig2}
\end{figure}

The essential goal of both analyses is to select events consistent with DIF 
candidate electrons, while eliminating remaining backgrounds from cosmic-ray 
interactions, including neutrons and photons. 
The electron identification relies primarily on the differences in the timing 
characteristics of the components of light produced in the event: scintillation 
light, and \v{C}erenkov light, both direct and rescattered - 
Figs.~\ref{Fig2}(a) and (b). 
Furthermore, the event likelihood fitting returns also the fraction of 
direct \v{Cerenkov} light in the event, which provides excellent rejection 
against neutrons - Fig.~\ref{Fig2}(c). 

High energy $\gamma$ rays, from $\pi^0$ produced by neutron interactions in
the lead shielding of the veto system, enter the detector fiducial volume
without leaving a veto signal. 
The charged particles resulting from their interactions in the liquid point 
predominantly into the detector volume and are difficult to distinguish from 
electrons from the $\nu_e \, C \to e^- \, X$ reaction on the basis of electron 
PID alone. 
The backwards projected track-length to the edge of the detector volume, $s$, 
is used to remove these events, concentrated at low values of $s$ - 
Fig.~\ref{Fig2}(d). 
Events with any veto hits in time with the event, and along the backward 
extrapolation of the track, are also rejected.

Finally, the electron events in the final DIF sample are required to have 
$\cos\theta_e < 0.8$, where $\theta_e$ is the angle between the reconstructed 
direction and the incident neutrino beam. 
This greatly reduces the BRB from the forward-peaked $\nu_\mu e$ elastic 
scattering, while retaining a high efficiency for the DIF signal. 

After applying all of the respective selection criteria, both analyses obtain 
a significant and consistent beam-related event excess. 
One analysis ends up with 23 beam-on events and 114 beam-off events 
(8.0 rescaled for the duty factor), which corresponds to 15.0 excess events. 
The other analysis ends up with 25 beam-on events and 92 beam-off events 
(6.4 rescaled), which corresponds to 18.6 excess events. 
Their efficiencies are 8.4\% and 13.8\%, respectively, calculated for 
to the $d>0$ fiducial volume.

%==============================================================================
%
%       5 - Background Processes
%

As already mentioned, the main BRBs in the DIF oscillation search come from 
the intrinsic $\nu_e$ contamination in the beam, 
$\mu^+ \to e^+ \nu_e \bar\nu_\mu$ (DIF), and $\pi^+ \to e^+ \nu_e$ (DIF). 
These backgrounds are calculated using the beam MC neutrino fluxes and the 
$\nu_e C$ cross section calculated in the CRPA model. 
The $\nu_\mu e$ elastic scattering background from the $\nu_\mu$ 
DIF flux is greatly reduced by requiring $\cos\theta_e < 0.8$. 
The last relevant background, $\pi^+ \to \mu^+ \nu_\mu$ DIF followed by 
$\nu_\mu C \to \nu_\mu C \pi^{\circ}$ coherent scattering, 
is calculated using the cross section in Ref.~\cite{Coh-pi0}. 
Backgrounds from the $\nu_\mu C \to \mu^- X$ reaction are negligible. 
The four relevant BRBs are summarized in Table~\ref{TableI}. 
The total BRBs calculated for the two analyses yield 4.5 and 8.5 events, 
respectively, which thus leaves a significant excess of events (10.5 and 10.1 
events, respectively) above the expectation from conventional processes. 
The probabilities that the number of expected background events (12.5/14.9) 
fluctuate up to the observed beam-on numbers (23/25) are 
$0.7\times10^{-2}$ and $1.6 \times 10^{-2}$, respectively.

Since both analyses have low efficiencies, different reconstruction software, 
and different selection criteria, the two samples need not necessarily be 
identical. 
Both the logical AND and OR of the two samples have been extensively studied 
in MC simulations and the results are consistent with the expectations. 
For the final DIF sample we have elected to use the logical OR of the events. 
This procedure minimizes the sensitivity of the measurement to uncertainties 
in the efficiency calculations, is less sensitive to statistical fluctuations, 
and also yields a larger overall efficiency. 
Table~\ref{TableII} summarizes the final event samples for the individual 
analyses, for their overlap (AND), and for the final sample (OR).

%==============================================================================
%
%       6 - Interpretation
%

\begin{figure}[h]
\begin{center}
\mbox{
\epsfxsize=0.9\columnwidth \epsfbox[20  20 550 550]{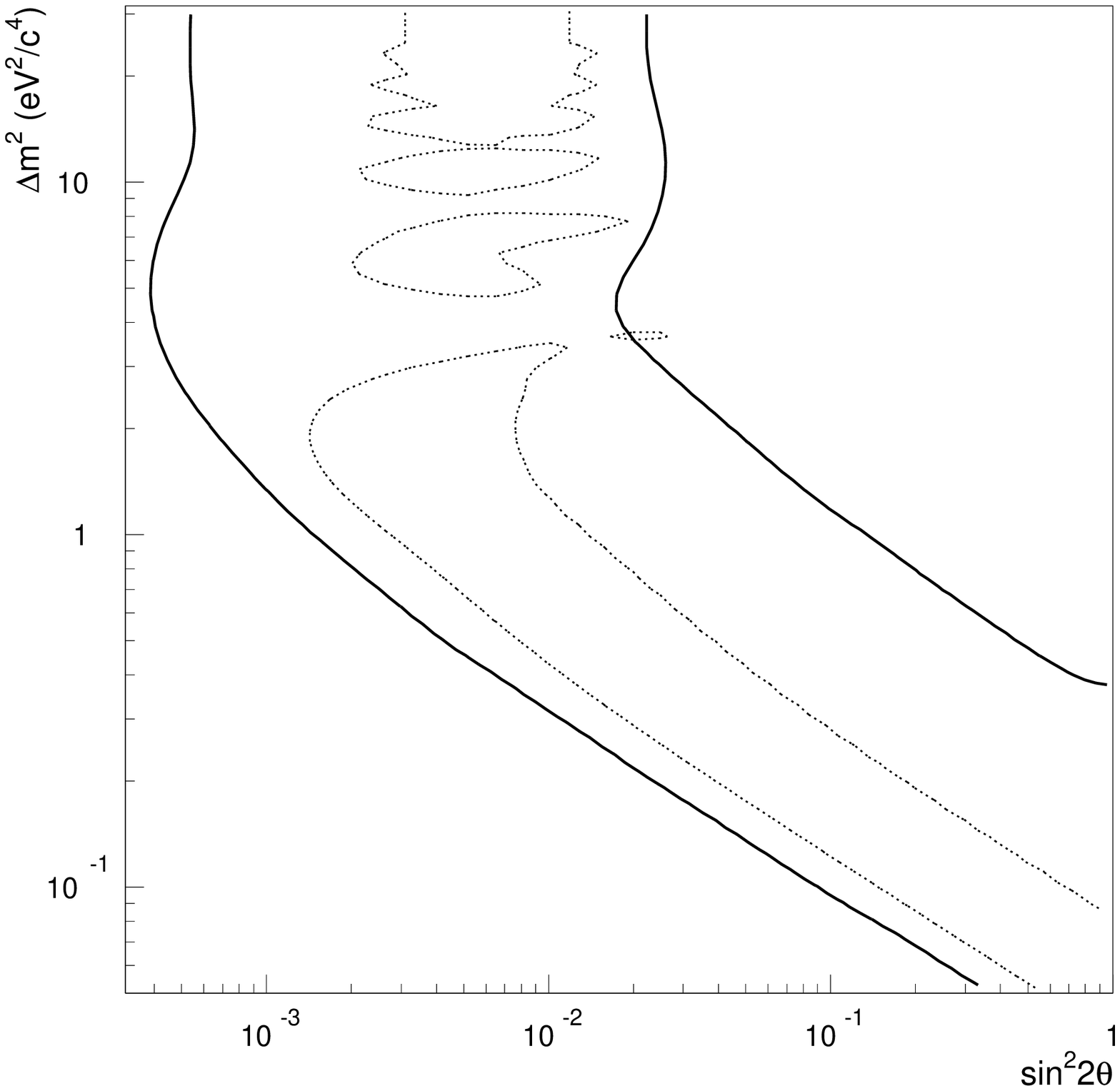}
}
\end{center}
\caption{
The 95\% confidence level region for the DIF $\nu_\mu \to \nu_e$ along with the 
favored regions from the LSND $\bar\nu_\mu \to \bar\nu_e$ DAR measurement 
(dotted contours). 
}
\label{Fig3}
\end{figure}

In the following we interpret the observed event excess of the OR sample in 
terms of the simplest, two-generation mixing neutrino oscillations formalism. 
In this model the oscillation probability is given by
\begin{equation}
P=\sin^2 2\theta \ \sin^2 \left(1.27 \ \Delta m^{2}{L\over E_\nu}\right),
\end{equation}
where $\theta$ is the mixing angle, $\Delta m^2$ (eV$^2$/c$^4$) is the 
difference of the squares of the masses of the appropriate mass eigenstates, 
$L$ (m) is the distance from neutrino production to detection, and $E_\nu$ 
(MeV) is the neutrino energy. 
Since the distance to the source is ambiguous because of the presence of 
multiple beam targets (A1, A2, and A6), the energy distribution alone is used 
to determine the confidence levels in the $(\sin^2 2\theta,\Delta m^2)$ 
parameter space. 
Fig.~\ref{Fig3} shows the 95\% confidence level contours that result from the 
fit. 
This result is consistent with the previous LSND DAR result~\cite{DAR-LSND}, 
shown superimposed in Fig.~\ref{Fig3}.
The oscillation probability is $(2.6 \pm 1.0 \pm 0.5) \times 10^{-3}$, where 
the second error is systematic, as described below.

The neutrino cross sections and fluxes constitute the largest source of 
systematic uncertainty for the DIF analysis.  
Although our measurement of $\nu_e C$ scattering using the DAR $\nu_e$ flux 
agrees well with calculations~\cite{nueC}, our measurement of the inclusive 
$\nu_\mu C$ cross section is 45\% below the CRPA calculation~\cite{numuC}. 
The $\nu_\mu$ flux in the $\nu_\mu C$ measurement is the same as for this DIF 
oscillation analysis, and it is possible that the $\nu_e C$ cross section at 
these higher energies also is below the CRPA calculation. 
The expected average number of events is given by
\begin{equation}
N_{total} =
\varepsilon \, \sigma_{\nu_e C} \,
(\Phi_{\nu_\mu}P_{\nu_\mu \to \nu_e} + \Phi_{\nu_e} ) + N_{BUB}
\end{equation}
where $\varepsilon$ is the event selection efficiency, $\Phi_{\nu_e/\nu_\mu}$
are the neutrino fluxes, $\sigma_{\nu_e C}$ is the neutrino cross section, and
$N_{BUB}$ is the beam-unrelated background (BUB). 
The oscillation signal is proportional to the same product 
($\varepsilon \, \sigma_{\nu_e C} \, \Phi_{\nu_\mu}$) as the neutrino 
background, since $\Phi_{\nu_e}$ is proportional to $\Phi_{\nu_\mu}$.
The effect of {\em lowering} the product 
$\varepsilon \, \Phi \, \sigma_{\nu_e C}$ 
is to reduce the predicted BRB, which raises the observed oscillation signal. 
Only by {\em raising} the product
$\varepsilon \, \Phi \, \sigma_{\nu_e C}$ 
is the oscillation signal decreased.
In order to calculate conservative confidence regions, we allow the 
value of $\varepsilon \, \Phi \, \sigma_{\nu_e C}$ to vary between
21\% above to 45\% below the calculated value. 
Only a symetrical 21\% systematic error is used in the oscillation probability.
%
%==============================================================================
%
%       Section 6:  Conclusions
%

We have described a search for $\nu_e \, C \to e^- \, X$ interactions for
electron energies $60 < E_e < 200$ MeV.
Two independent analyses observe a number of beam-on events significantly
above the expected number from the sum of conventional beam-related processes
and cosmic-ray (beam-off) events.
The probability that the $21.9 \pm 2.1$ estimated background events fluctuate 
into 40 observed events is $1.1\times 10^{-3}$. 
The excess events are consistent with $\nu_\mu \to \nu_e$ oscillations 
with an oscillation probability of $(2.6 \pm 1.0 \pm 0.5) \times 10^{-3}$. 
A fit to the energy distribution events, assuming neutrino oscillations as the 
source of $\nu_e$, yields the allowed region in the 
$(\sin^2 2\theta,\Delta m^2)$ parameter space shown in Fig.~\ref{Fig3}. 
This allowed region is consistent with the allowed region from the DAR search 
reported earlier by LSND.
This $\nu_\mu \to \nu_e$ DIF oscillation search has completely different 
backgrounds and systematic errors from the $\bar\nu_\mu \to \bar\nu_e$ DAR 
oscillation search and provides additional evidence that both effects are due 
to neutrino oscillations.

\paragraph*{Acknowledgments}
This work was conducted under the auspices of the US Department of Energy, 
supported in part by funds provided by the University of California for 
the conduct of discretionary research by Los Alamos National Laboratory. 
This work is also supported by the National Science Foundation. 

%==============================================================================
%
%	References
%

%==============================================================================
%
% Tables
%
%..............................................................................
\begin{table}
\caption{
Background estimates for the $\nu_\mu \to \nu_e$ oscillation search 
for the $d > 0$ fiducial volume, $9.2 \times 10^{22}$ POT, 
and for electron energies between 60 MeV and 200 MeV. 
These numbers are illustrative for an electron selection efficiency of 0.10, 
independent of energy. 
The actual efficiencies in the two analyses are slightly different and energy 
dependent.
}
\begin{tabular}{lcccc}
Process                               &          Flux         &
    $<\sigma>_\nu$      & Eff. & Number    \\
                                      & (cm$^{-2}$POT$^{-1}$) &
 ($10^{-40}$ cm$^{-2}$) & (\%) & of Events \\
\hline
$\nu_e   C \to e^- X$ ($\mu$ DIF)     & $3.8 \times 10^{-14}$ &
 28.3                   & 10.0 &    3.8    \\ 
$\nu_e   C \to e^- X$ ($\pi$ DIF)     & $8.3 \times 10^{-15}$ &
 79.2                   & 10.0 &    1.6    \\
$\nu_\mu C \to \nu_\mu C \pi^{\circ}$ & $6.5 \times 10^{-11}$ &
  1.6                   &  6.0 &    0.3    \\
$\nu_\mu e \to \nu_\mu e$             & $6.5 \times 10^{-11}$ &
  0.00136               &  0.5 &    0.1    \\
\hline
Total background                      &                       &
                        &      &    5.8    \\
\end{tabular}
\label{TableI}
\end{table}
%..............................................................................
\begin{table}
\caption{
Comparison of results for the two analyses (labeled here as A and B), their 
logical AND and OR. 
All errors are statistical.
}
\begin{tabular}{lcccccc}
Data &  Beam  &      BUB     &     BRB     &    Excess    & Eff. &
 Osc. Prob.         \\
Set  & On/Off &              &             &              & (\%) &
 $(\times 10^{-3})$ \\ 
\hline
A    & 23/114 & $ 8.0\pm0.7$ & $4.5\pm0.9$ & $10.5\pm4.9$ &  8.4 &
 $2.9\pm1.4$ \\ 
B    & 25/ 92 & $ 6.4\pm0.7$ & $8.5\pm1.7$ & $10.1\pm5.3$ & 13.8 &
 $1.7\pm0.9$ \\
AND  &  8/ 31 & $ 2.2\pm0.3$ & $3.1\pm0.6$ & $ 2.7\pm2.9$ &  5.5 &
 $1.1\pm1.2$ \\
OR   & 40/175 & $12.3\pm0.9$ & $9.6\pm1.9$ & $18.1\pm6.6$ & 16.5 &
 $2.6\pm1.0$ \\
\end{tabular}
\label{TableII}
\end{table}
%..............................................................................
%
%==============================================================================
%
% Figures
%..............................................................................
%
%..............................................................................
%
%
%..............................................................................
%
%..............................................................................
%

\begin{references}

\bibitem{DAR-LSND} C.\ Athanassopoulos {\it et.\ al.\ } (LSND Collaboration),
      Phys.\ Rev.\ C {\bf 54}, 2685 (1996); 
      C.\ Athanassopoulos {\it et.\ al.\ } (LSND Collaboration),
      Phys.\ Rev.\ Lett.\ {\bf 77}, 3082 (1996).
\bibitem{Burman-MC} R.\ L.\ Burman, M.\ E.\ Potter, and E.\ S.\ Smith, 
      Nucl.\ Instrum.\ Methods A {\bf 291}, 621 (1990); 
      R.\ L.\ Burman, A.\ C.\ Dodd, and P.\ Plischke,
      Nucl.\ Instrum.\ Methods in Phys.\ Research A {\bf 368}, 416 (1996).
\bibitem{numuC} C.\ Athanassopoulos  {\it et.\ al.\ } (LSND Collaboration), 
      LA-UR-97-1848, to appear in Phys.\ Rev.\ C.
\bibitem{NIM-LSND} C.\ Athanassopoulos  {\it et.\ al.\ } (LSND Collaboration), 
      Nucl.\ Instrum.\ Methods A {\bf 388}, 149 (1997).
\bibitem{E645} J.\ J.\ Napolitano {\it et.\ al.\ },
      Nucl.\ Instrum.\ Methods A {\bf 274}, 152 (1989).
\bibitem{CRPA} E.\ Kolbe, K.\ Langanke, and S.\ Krewald, 
      Phys.\ Rev.\ C {\bf 49}, 1122 (1994); 
      (K.\ Langanke, private communication).
\bibitem{DIF-LSND} C.\ Athanassopoulos  {\it et.\ al.\ } (LSND Collaboration), 
      LA-UR-97-1998/UCRHEP-E191, submitted to Phys.\ Rev.\ C.
\bibitem{Coh-pi0} D.\ Rein and L.\ M.\ Sehgal, 
      Nucl.\ Phys.\ B {\bf 223}, 29 (1983).
\bibitem{nueC} C.\ Athanassopoulos  {\it et.\ al.\ } (LSND Collaboration), 
      Phys.\ Rev.\ C {\bf 55}, 2078 (1997).

\end{references}
\end{document}